\documentclass{appolb}
\usepackage{epsfig,graphicx,amsmath,dsfont}

\begin{document}
\title{Initial conditions, equations of state and final state in hydrodynamics
\thanks{Presented at the IV Workshop on Particle Correlations and Femtoscopy}}
\author{M\'at\'e Csan\'ad
\address{E{\"o}tv{\"o}s Lor{\'a}nd University, H - 1117 Budapest,
P{\'a}zm{\'a}ny P. s. 1/A, Hungary}
}
\maketitle
\begin{abstract}
In this paper we present properties of relativistic and non-relativistic perfect hydrodynamical models. In particular we show illustrations of the fact that different initial conditions and equations of state can lead to the same hadronic final state. This means that alone from the hadronic observables one cannot determine either of the above, one needs for example penetrating probes that inherit their properties from each timeslice of the evolution of the fireball.
\end{abstract}


\section{Perfect fluid hydrodynamics}
Perfect fluid hydrodynamics is based on local conservation of
entropy or density, energy and momentum, expressed by so-called conservation equations. The fluid is perfect if the energy-momentum tensor is diagonal in the local rest frame, i.e. there are no shear stress, viscosity or heat conduction effects. The conservation equations are closed by the equation of state, which gives the relationship between energy density $\epsilon$, pressure $p$. Typically $\epsilon = \kappa p$, where the proportionality ``constant'' $\kappa$ may depend on temperature $T$, which is connected to the density $n$ and pressure $p$ via $p=nT$ in many solutions of the exactly solvable class, especially those discussed in following. In addition a bag constant $B$ can be introduced with a modified equation of state of $\epsilon - B = \kappa (p+B)$, where $B=0$ in the hadronic phase, non-zero in a deconfined phase. In this paper only the $\epsilon = \kappa p$ case is discussed.

Solutions describing non-relativistic flows that are applicable to relativistic heavy ion collisions are e.g. described in refs.~\cite{Csorgo:2001xm,Csorgo:2003rt,Sinyukov:2004am}, we will investigate a family of such solutions.

Solving the relativistic equations is much harder. There are only a few exact solutions for these equations. One (and historically the first) is the Landau-Khalatnikov implicit solution solution discovered more than 50 years
ago~\cite{Landau:1953gs,Khalatnikov:1954aa,Belenkij:1956cd}. This
is a 1+1 dimensional solution, and has realistic properties: it
describes a 1+1 dimensional expansion, does not lack acceleration
and predicts an approximately Gaussian rapidity distribution.

Another renowned solution of relativistic hydrodynamics is the
Hwa-Bjorken solution~\cite{Hwa:1974gn,Chiu:1975hw,Bjorken:1982qr},
which is a simple, 1+1 dimensional, explicit and exact, but accelerationless solution. This solution is boost-invariant in its original form, hence fails to describe the data~\cite{Back:2001bq,Bearden:2001qq}. However, the solution
allowed Bjorken to obtain a simple estimate of the initial energy
density reached in high energy reactions from final state hadronic
observables.

Important are solutions~\cite{Csorgo:2006ax,Bialas:2007iu} which are explicit and describe a relativistic acceleration, i.e. combine the properties of the above solutions. We will investigate such a family of solutions.

\section{Investigated solutions}\label{s:sols}

We will investigate here an exact relativistic and an exact non-relativistic solution. We will use these solutions to extract information on the dependence of the final state on initial state, the parameters of the exact solution and the equation of state used.

\subsubsection*{A non-relativistic hydro solution}

The below discussed non-relativistic solution describes a 3+1 dimensional ellipsoidally symmetric expansion~\cite{Csorgo:2001xm,Csorgo:2001ru}, with an arbitrary $\kappa\in\mathds{R}$. The velocity field $\bf{v}$ is
\begin{equation}
{\bf v}=\left(\frac{\dot X}{X}r_x, \frac{\dot Y}{Y}r_y, \frac{\dot Z}{Z} r_z\right),
\end{equation}
where $X$, $Y$ and $Z$ are time-dependent principal axes of the expanding ellipsoid, $\dot X$, $\dot Y$ and $\dot Z$ are their expansion rate versus time, and $r=(r_x,r_y,r_z)$ are the spatial coordinates. As for the thermodynamical quantities, one has
\begin{gather}
n(r,t)=n_0\frac{X_0 Y_0 Z_0}{XYZ} \exp\left(-\frac{r_x^2}{2X^2}-\frac{r_y^2}{2Y^2}-\frac{r_z^2}{2Z^2}\right),\\
T(r,t)=T_0\left(\frac{X_0 Y_0 Z_0}{XYZ}\right)^{1/\kappa},
\end{gather}
where $X_0=X(t_0)$, $Y_0=Y(t_0)$, $Z_0=Z(t_0)$ are the principal axes at a given (arbitrarily chosen) time $t_0$, and $n_0 = n(0,t_0)$,  $T_0=T(0,t_0)$. This represents a solution if the ordinary differential equations
\begin{equation}
\ddot X X=\ddot Y Y=\ddot Z Z= \alpha\left(\frac{X_0 Y_0 Z_0}{XYZ}\right)^{1/\kappa}
\end{equation}
are fulfilled. Here $\alpha$ is an ``acceleration parameter'', in the above mentioned solution $\alpha=T_0/m$, but note, that with a generalization in the expression for $n(r,t)$ $\alpha$ can be made arbitrary.

\subsubsection*{A relativistic hydro solution}

Next being discussed is a family of analytic, explicit and simple solutions, which do not lack acceleration, and yield finite, realistic rapidity distributions~\cite{Csorgo:2006ax}. In these solutions the velocity field is given by
\begin{equation}
u^\mu=\gamma(1,\bf{v})
\end{equation}
with $\bf{v}$ being the $d$ dimensional spherically symmetric velocity. The length of this vector is denoted by $v$, and this length can be expressed in the Rindler coordinates $\eta$ and $\tau$ by
\begin{equation}
v=\tanh(\lambda\eta)
\end{equation}
where $\lambda$ is a kind of ``acceleration parameter'', because relativistic acceleration $u_\mu\partial^\mu u^\nu$ vanishes if $\lambda=1$, but is non-vanishing for a positive $\lambda$ different from 1. The pressure $p$ and the temperature $T$ are given by
\begin{gather}
p=p_0
\left(\frac{\tau_0}{\tau}\right)^{\lambda d\frac{\kappa+1}{\kappa}},\\
T=T_0
\left(\frac{\tau_0}{\tau}\right)^{\lambda d\frac{1}{\kappa}}.
\end{gather}
The value of the constants $\lambda$ (``acceleration parameter''), $d$ (number of spatial dimensions) and $\kappa$ (adiabatic index of the fluid) are constrained, and different possible set of values yield different solutions. For example if $\lambda=1$, then $d\in\mathds{R}$ and $\kappa\in\mathds{R}$ are arbitrary (this is corresponds to the Bjorken solution). In this case there is no acceleration. If however $\lambda=2$, i.e. there is substantial acceleration, $d\in\mathds{R}$ is arbitrary but $\kappa=d$ must be fulfilled. Another important case is, that if $d=1$ and $\kappa=1$, then $\lambda\in\mathds{R}$ is arbitrary. This last case has a remarkably general velocity field: the $\lambda$ acceleration parameter can be arbitrary. On the other hand, this solution works only for $d=1$ and $\kappa=1$, which is obviously a drawback. For a discussion of greater detail  Refs.~\cite{Csorgo:2006ax,Nagy:2007xn} are recommended.

\section{Dependence on the initial conditions and equation of state}
To calculate observables from the above solutions, freeze-out criteria have to be utilized. For example the simple condition of $T(r,t)=T_{\rm{freeze-out}}$ can be used, or a condition where the freeze-out hypersurface is pseudo-orthogonal to the velocity field. This is an important aspect of hydrodynamics, that a freeze-out condition has to be chosen, but here it does not need to be discussed: we fix the hydrodynamical final state and investigate only those hydrodynamical evolutions that lead exactly to the same final state (i.e. to the same hadronic observables).

\begin{figure}
\includegraphics[width=0.495\linewidth]{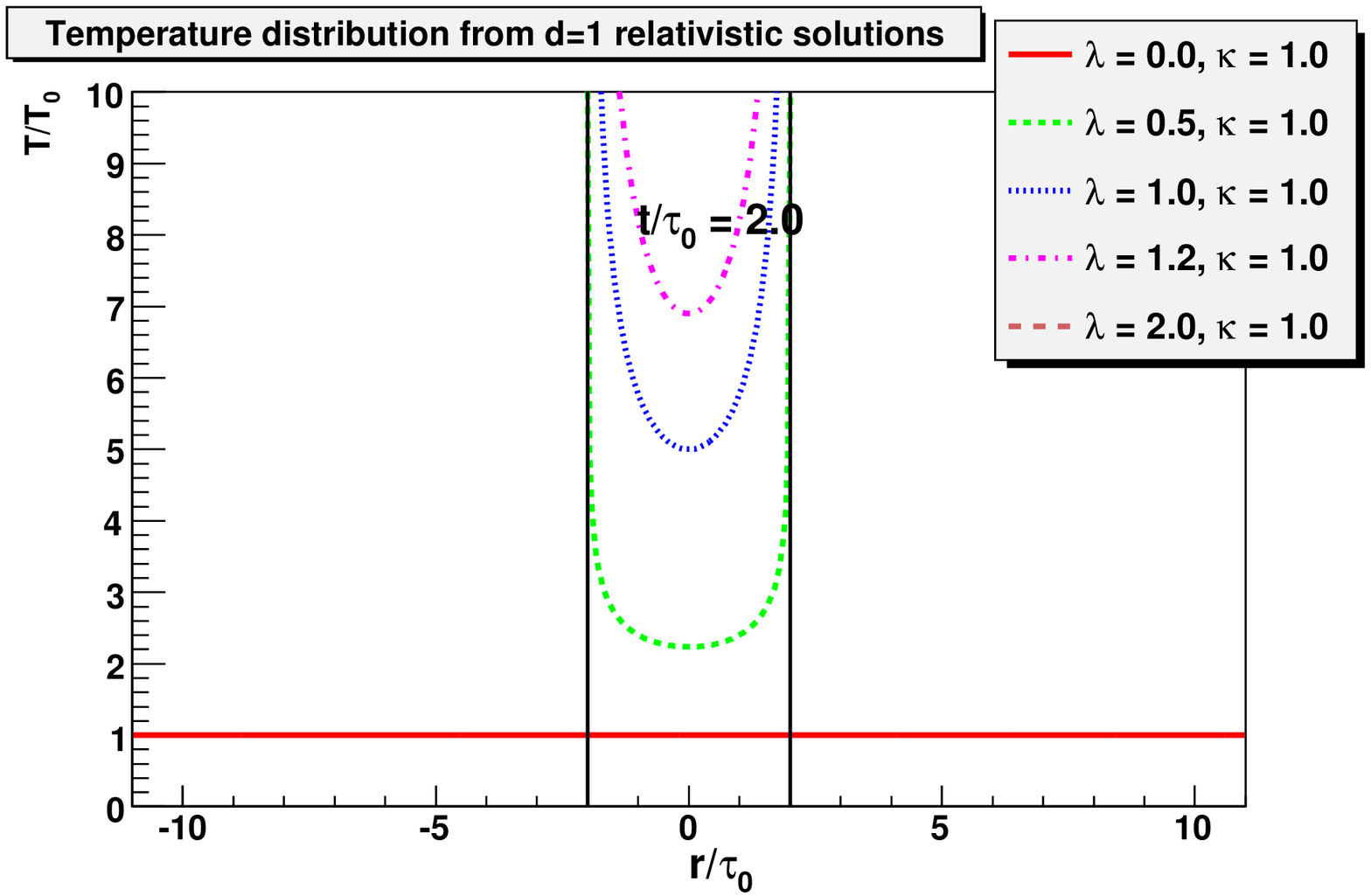} \includegraphics[width=0.495\linewidth]{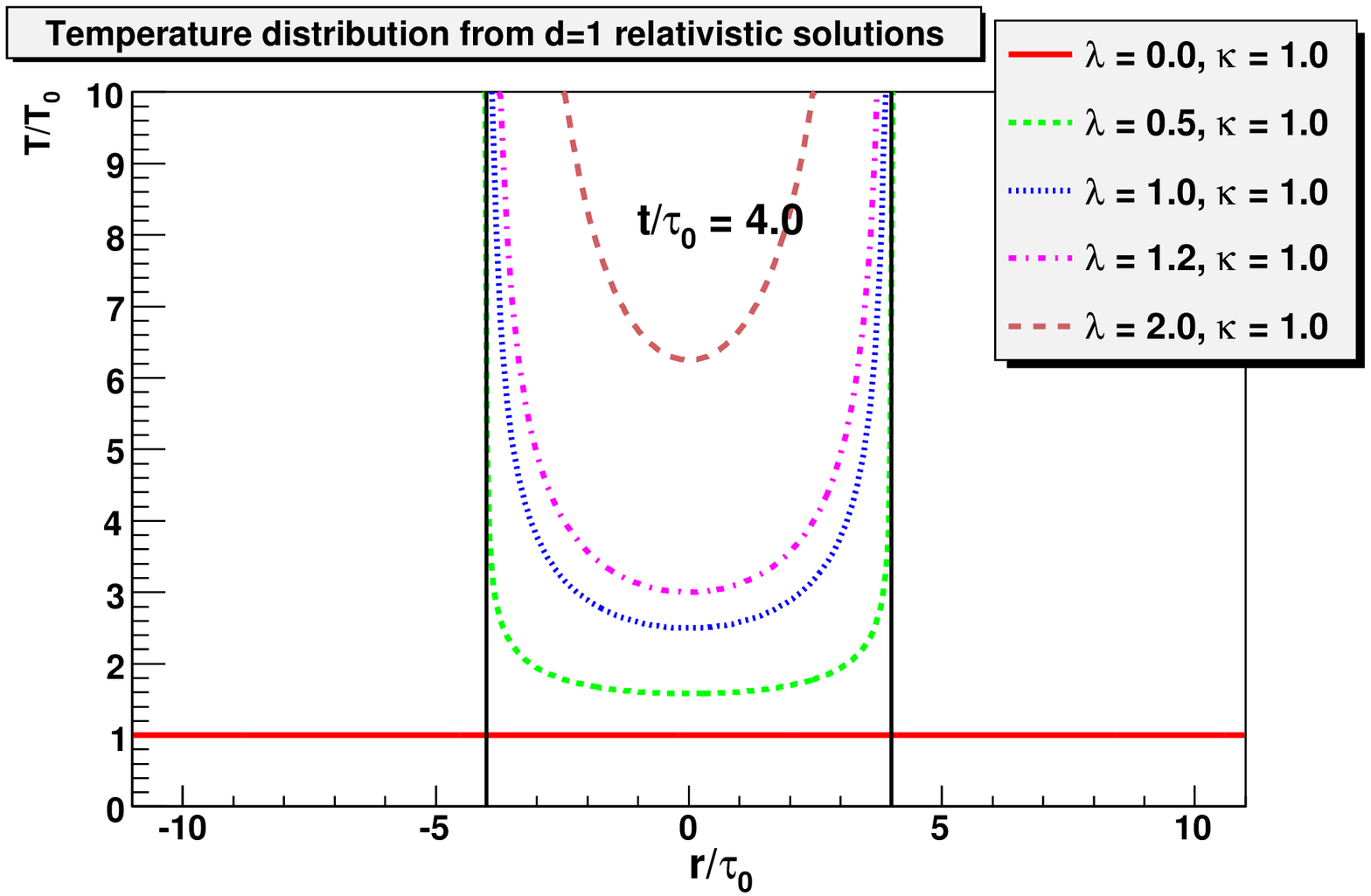}\\
\includegraphics[width=0.495\linewidth]{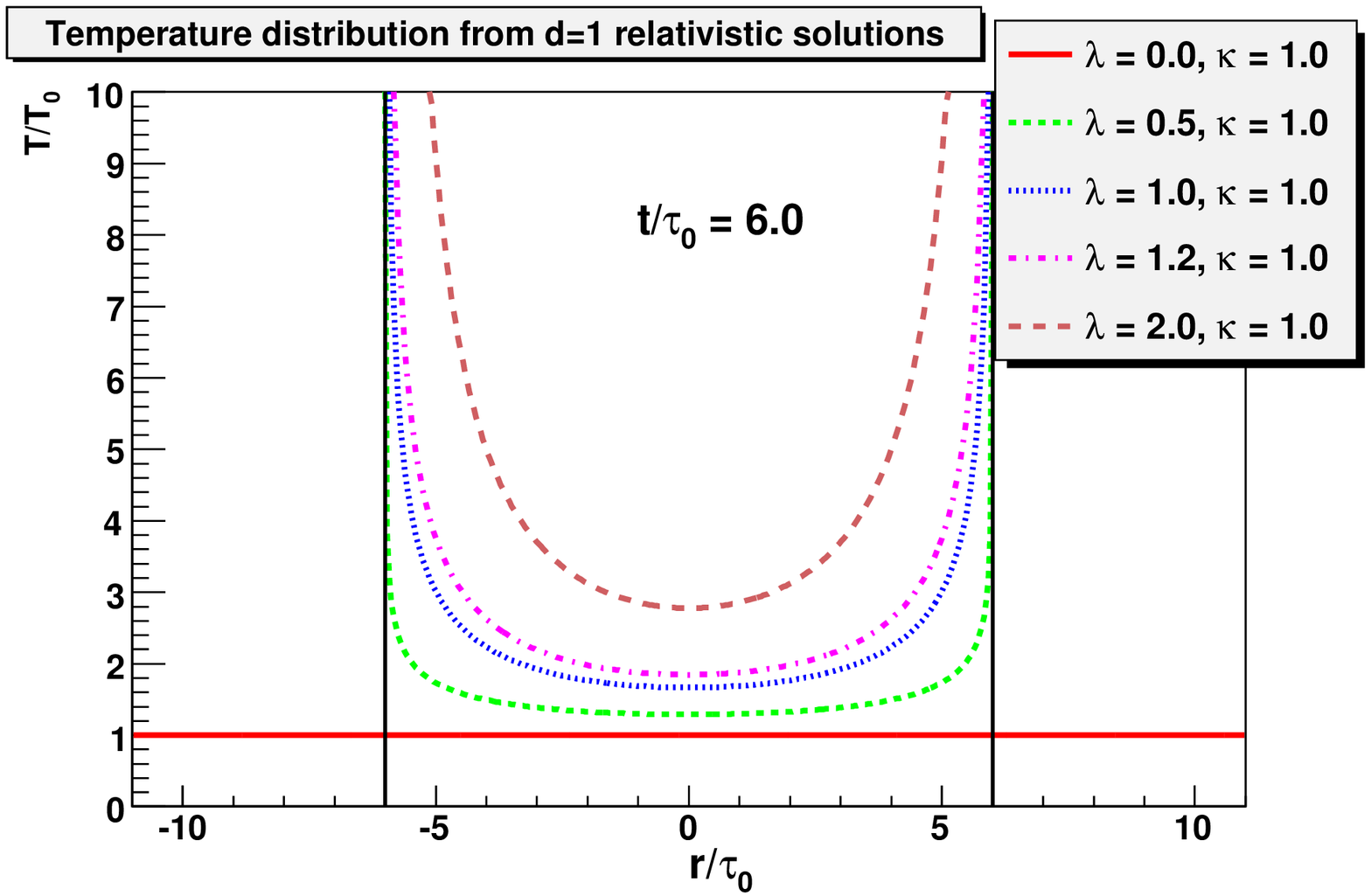}
\includegraphics[width=0.495\linewidth]{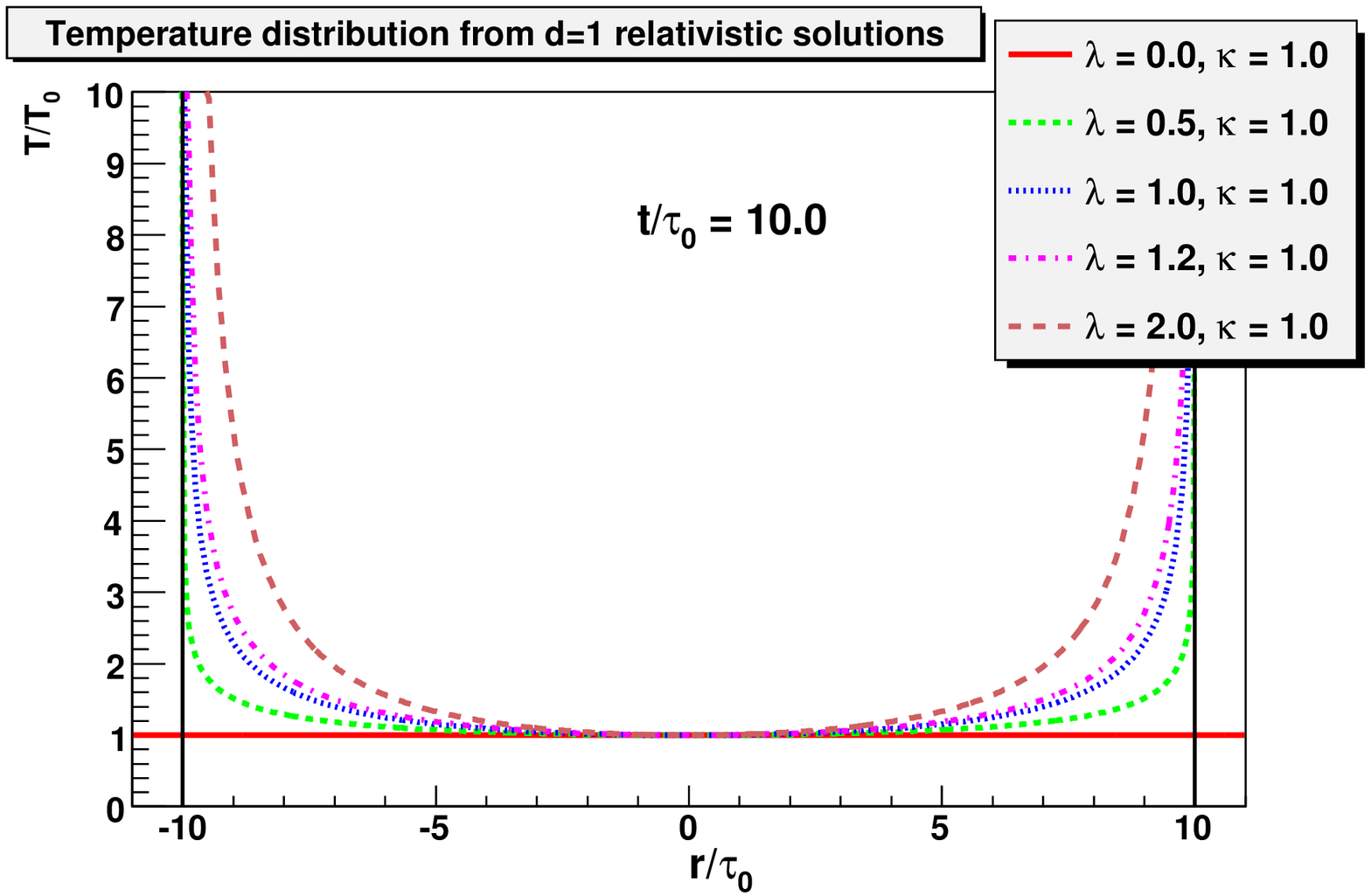}
\caption{\label{f:anim1}These plots show the
temperature distribution of a $d=1$ exact relativistic solutions from section~\ref{s:sols}, for different $\lambda$ parameters and different initial temperature distributions. All approximate the same final temperature distribution ($T\approx T_0$).}
\end{figure}

\begin{figure}
\includegraphics[width=0.495\linewidth]{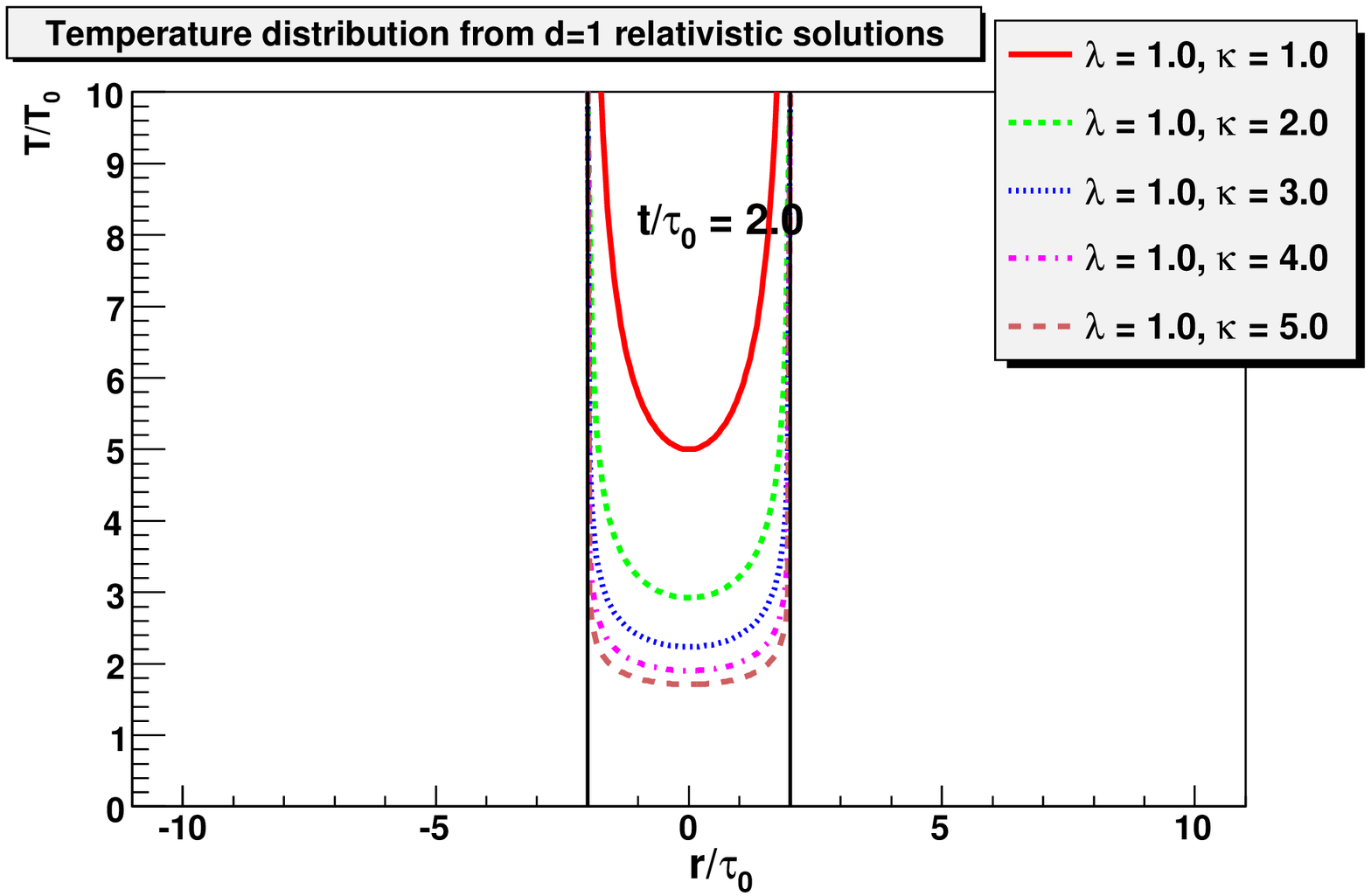} \includegraphics[width=0.495\linewidth]{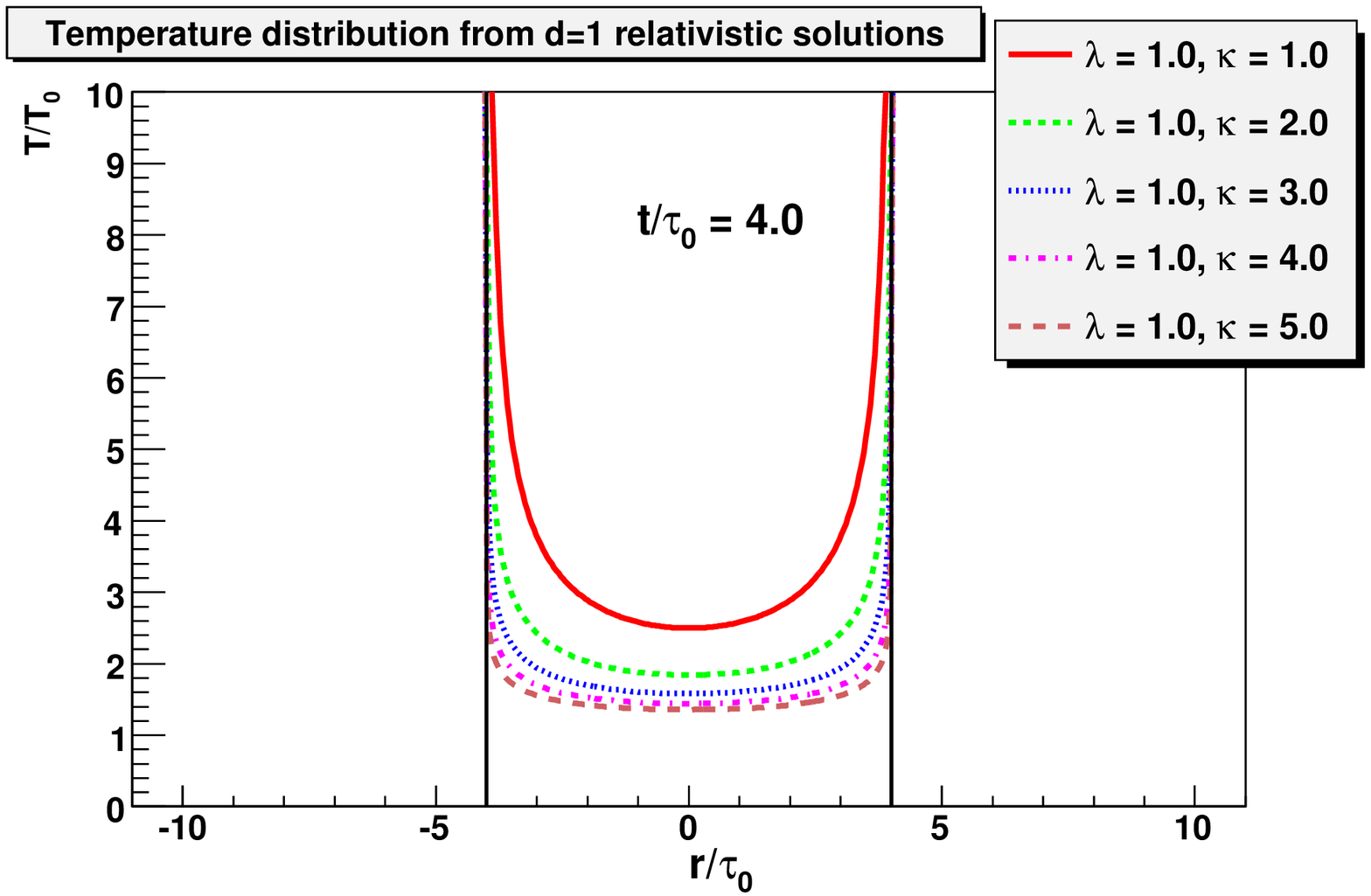}\\
\includegraphics[width=0.495\linewidth]{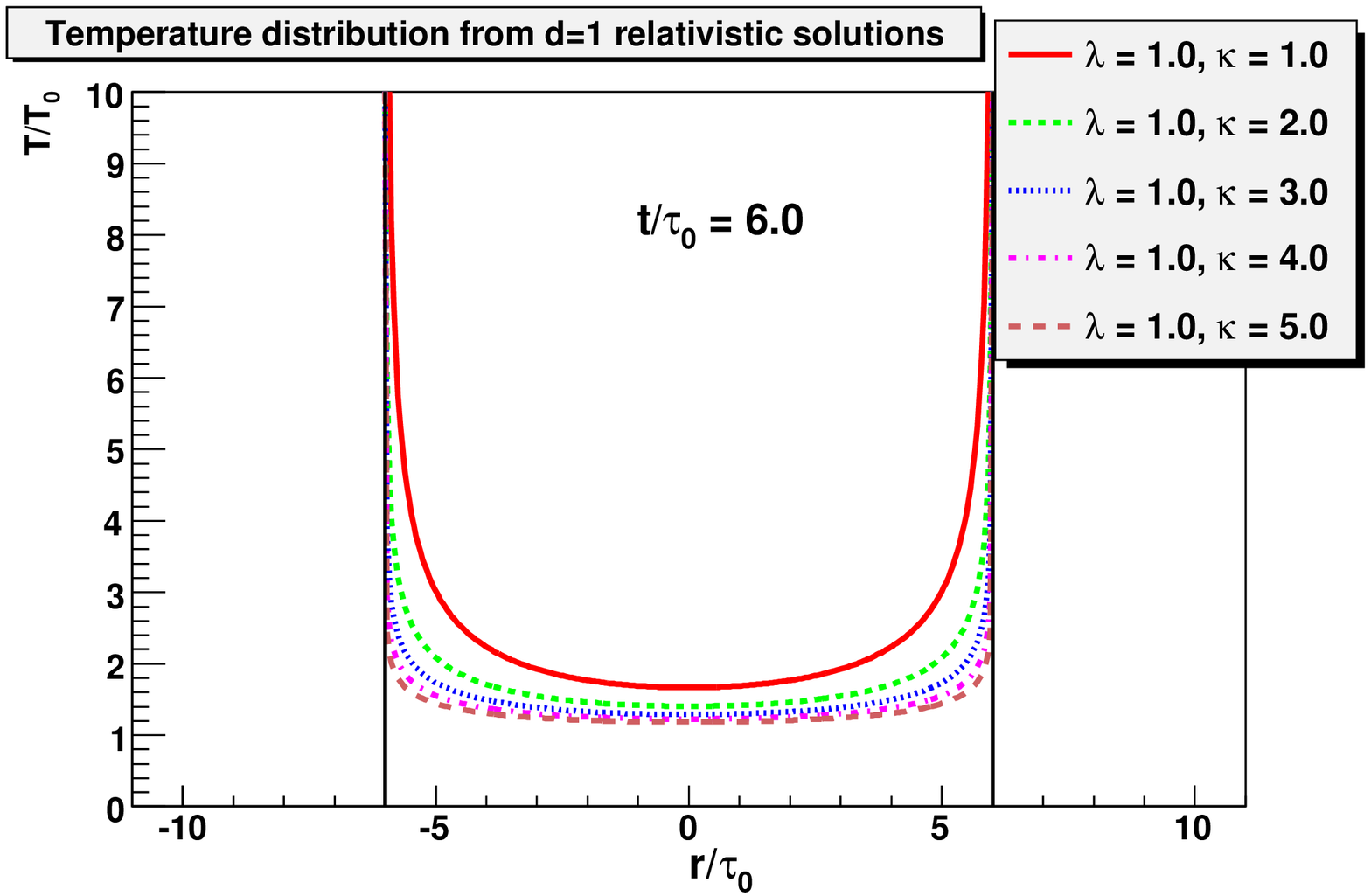}
\includegraphics[width=0.495\linewidth]{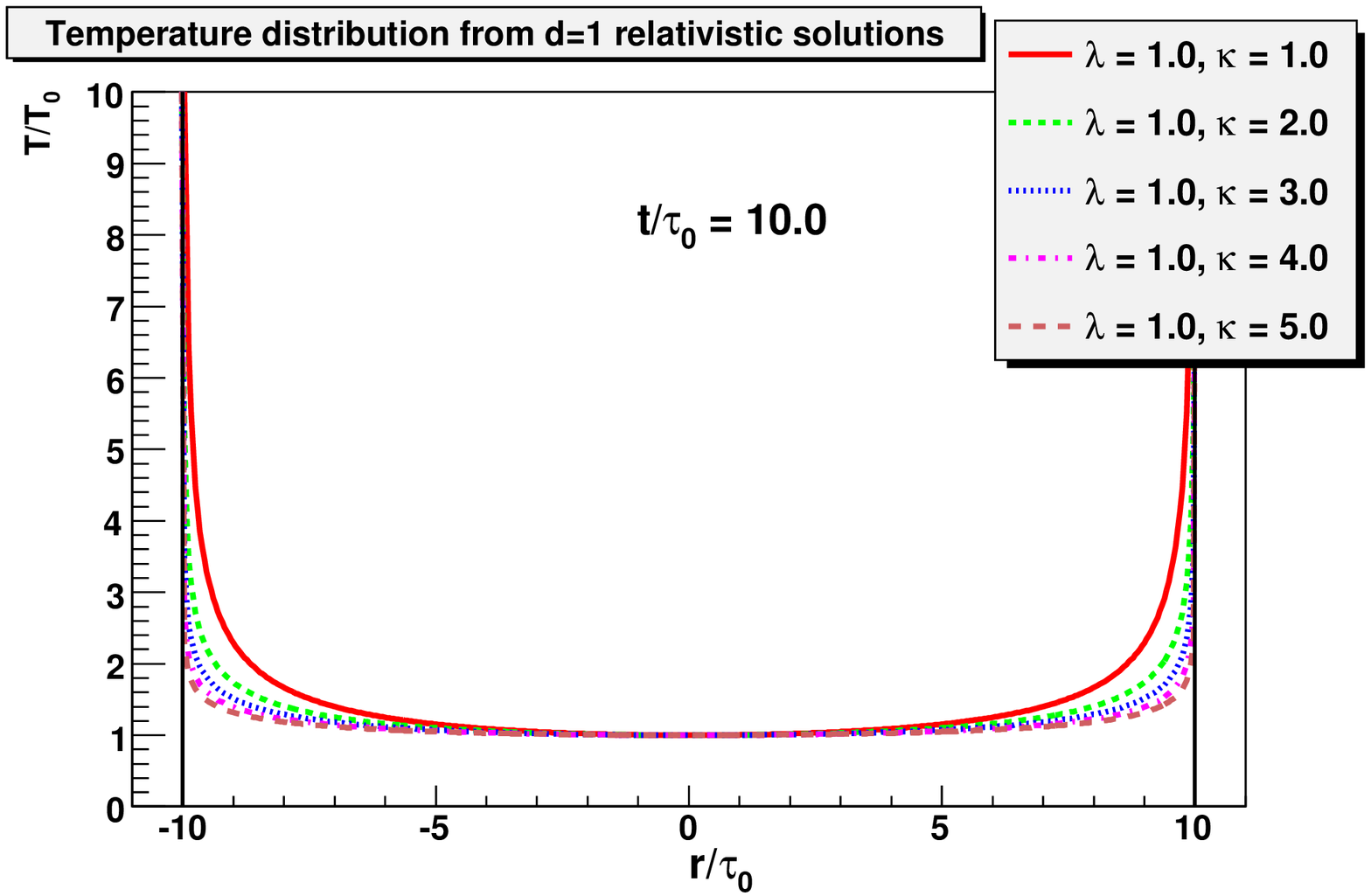}
\caption{\label{f:anim2}Temperature distribution is shown for a $d=1$ exact relativistic solution from section~\ref{s:sols} with different initial temperatures and equations of state ($\kappa$). All approximate the same final temperature distribution.}
\end{figure}

First, let us see relativistic solutions of section~\ref{s:sols}. In Figs.~\ref{f:anim1}~and~\ref{f:anim2} the spatial temperature distribution is shown at different times, for different initial conditions and different $\lambda$ and/or $\kappa$ values. All the plotted solutions go to the same final temperature at mid-rapidity
($\eta=0$), although their initial condition, acceleration
parameter and/or equation of state is different.

\begin{figure}
  \begin{center}
  \includegraphics[height=0.325\linewidth,angle=-90]{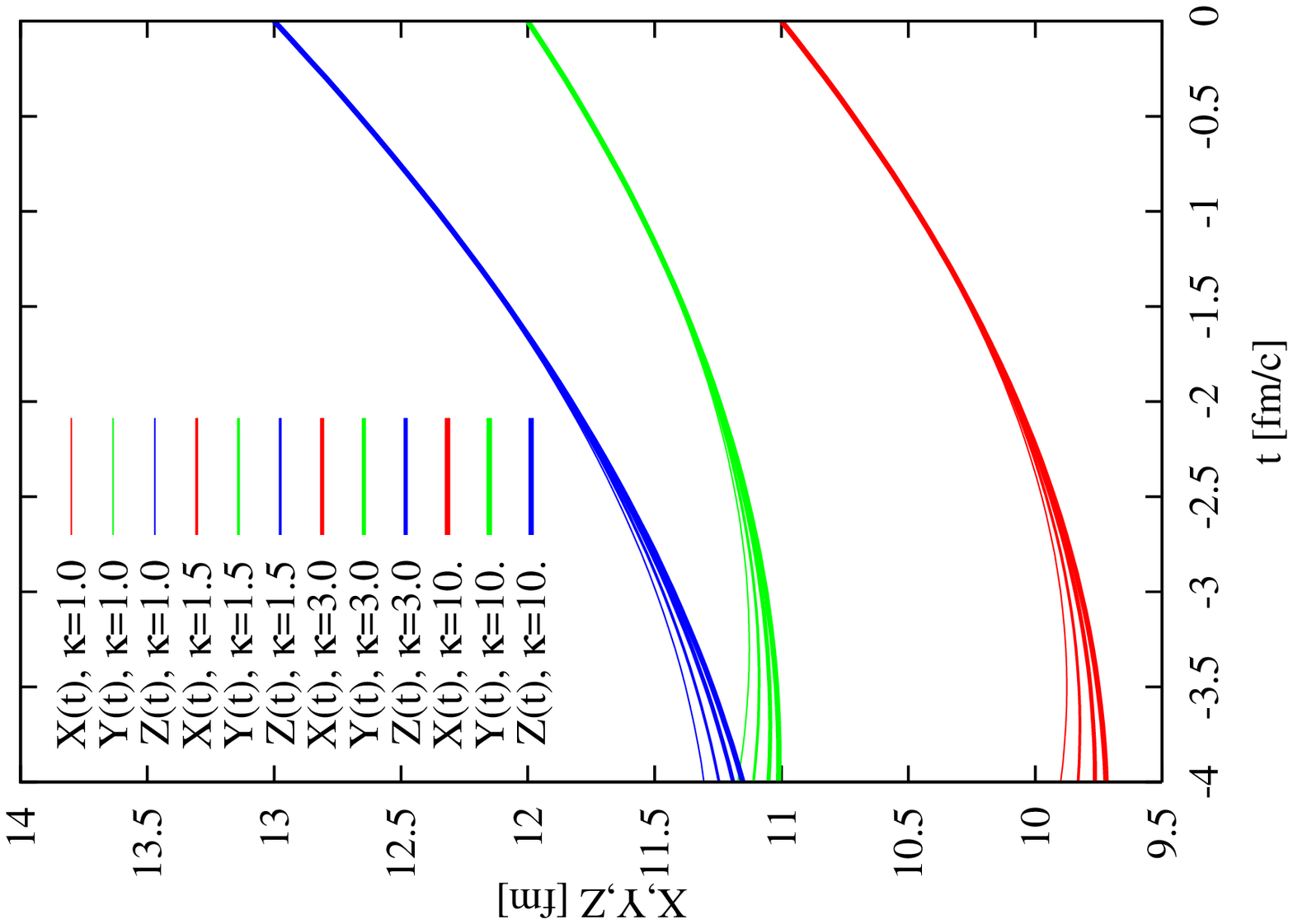}
  \includegraphics[height=0.325\linewidth,angle=-90]{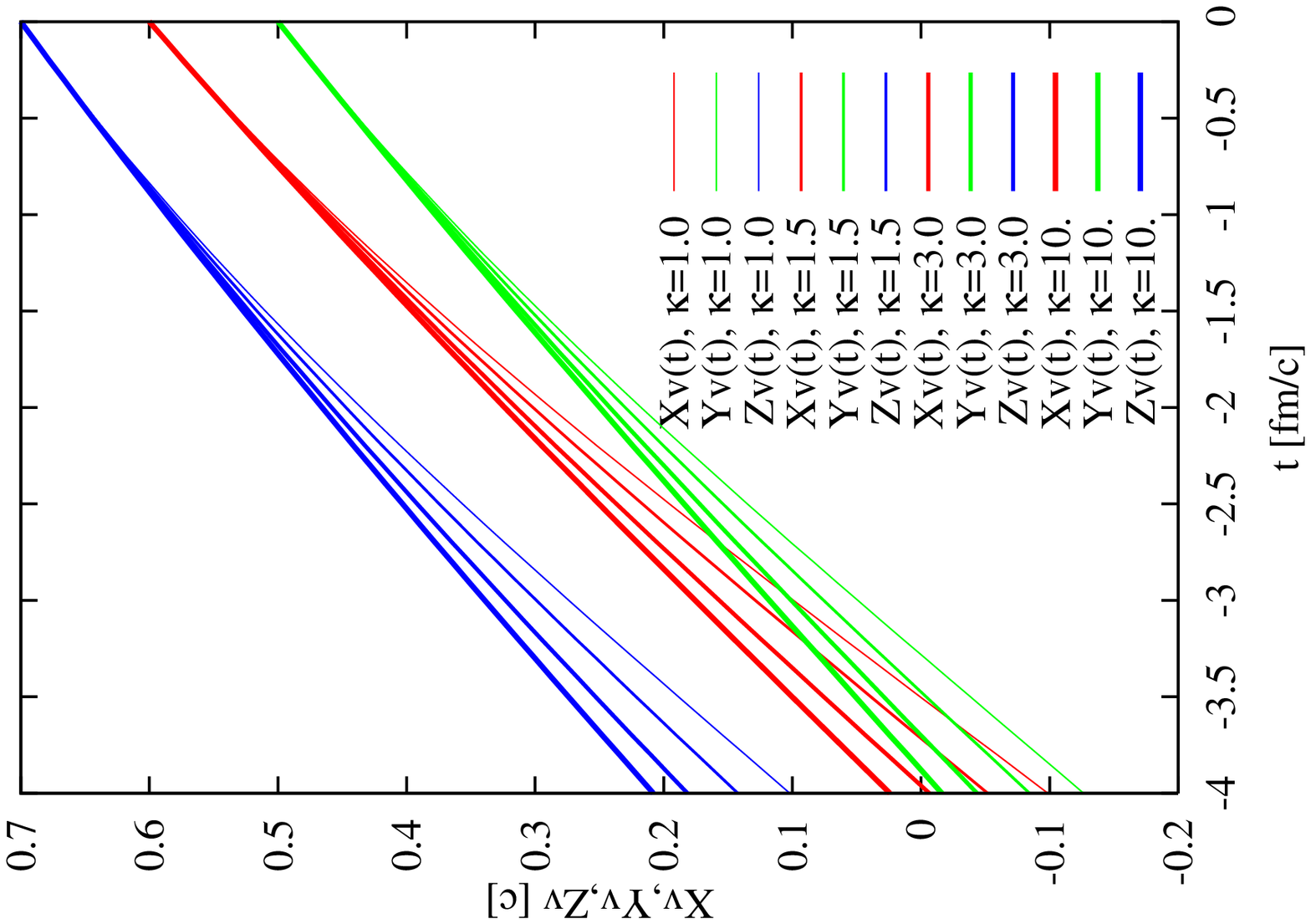}
  \includegraphics[height=0.325\linewidth,angle=-90]{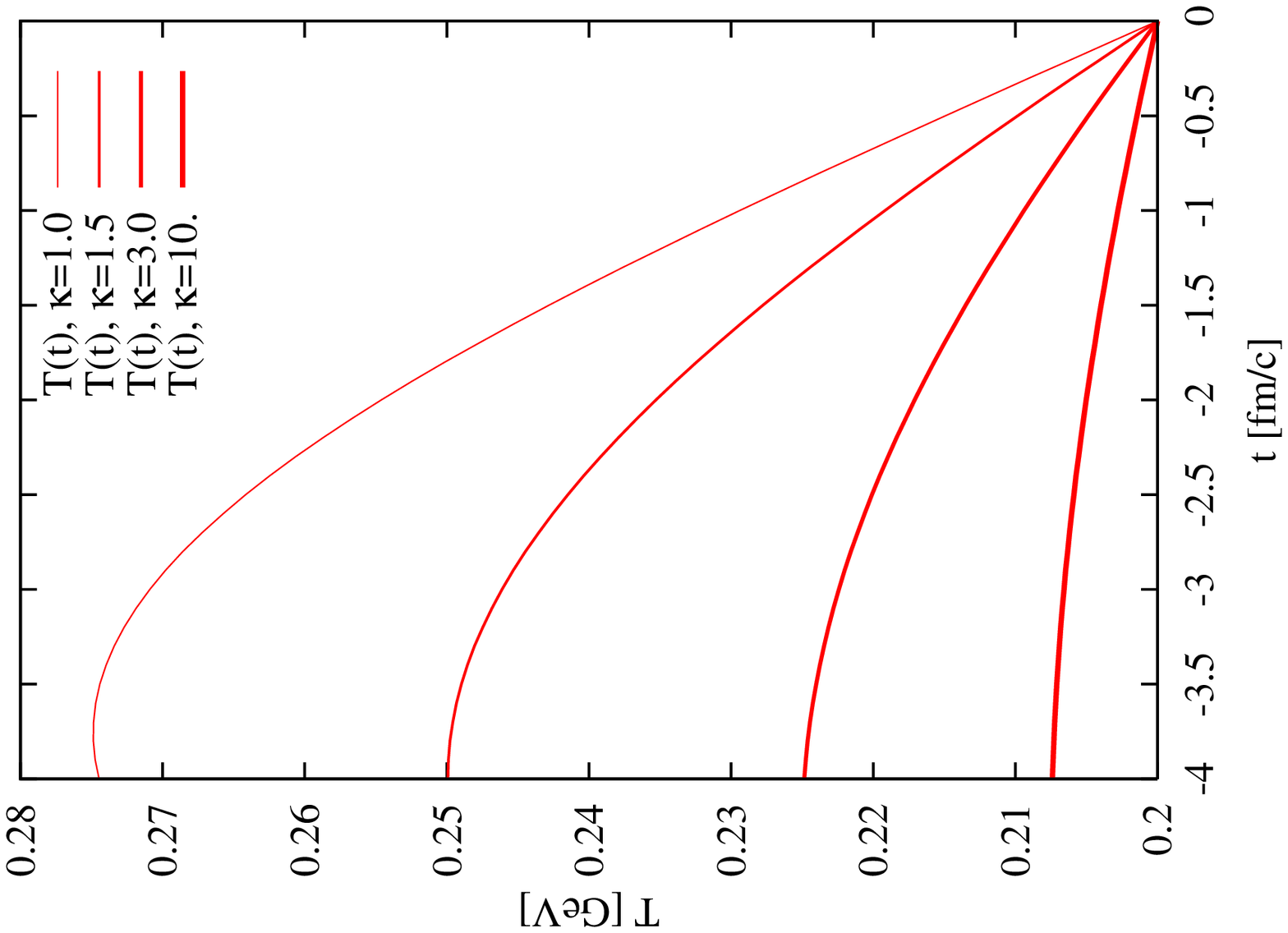}
  \end{center}
\caption{\label{f:timeev}Time evolution of the principal axes of the expanding ellipsoid ($X,Y,Z$), their expansion rates ($\dot X,\dot Y, \dot Z$, denoted by $\textrm{Xv}$ etc.), as well as that of the temperature ($T$) is plotted, for different equations of state ($\kappa$). It is evident that different equations of states combined with different initial conditions can lead to the same final state. The solution used here is the exact non-relativistic solution of section~\ref{s:sols}.}
\end{figure}

Then let us see non-relativistic solutions of section~\ref{s:sols}. In Fig.~\ref{f:timeev} time evolution of the principal axes of the expanding ellipsoid, their expansion rates and the temperature is shown, for different equations of state and different initial conditions. Here parameters were chosen so that the final states of all cases are the same: $T_0 = 200$ MeV, $X= 11$ fm, $Y= 12$ fm, $Z = 13$ fm, $\dot X_0 = 0.6$, $\dot Y_0 = 0.5$, $\dot Z_0 = 0.7$.

\section*{Conclusions}
We have seen that the same hydrodynamical final state can be achieved with different solutions, equations of state or initial conditions A very important consequence is also, that the hydrodynamic scaling laws observed in relativistic heavy ion collisions~\cite{Csanad:2005gv} do not depend on the initial conditions or the equation of state separately - one needs to restrict one in order to extract information on the other. These examples show that in order to extract the initial conditions or the equation of state from hadronic (final-state) observables, one needs further restrictions or experimental constraints. With penetrating probes one has access to the earlier times of the evolution thus can restrict results on initial conditions or equation of state.


\subsubsection*{Acknowledgements}
The author wishes to thank Tam\'as Cs\"org\H{o} and M\'arton Nagy for their valuable discussions, as well as the organizers of the IV Workshop on Particle Correlations and Femtoscopy for their kind hospitality.
This research was supported by the OTKA grants NK73143 and T049466, as well as the exchange program of the Hungarian Academy of Sciences and the Polish Academy of Arts and Sciences.

\bibliographystyle{prlsty}
\bibliography{Master}

\end{document}